\pdfoutput=1

\documentclass[11pt]{article}

\usepackage[preprint]{acl}

\usepackage{times}
\usepackage{latexsym}
 \usepackage{amsmath}
 \usepackage{booktabs}
 \usepackage{tabularx}
 \usepackage{amssymb}
 \usepackage{url}

\usepackage[T1]{fontenc}

\usepackage[utf8]{inputenc}

\usepackage{microtype}

\usepackage{inconsolata}

\usepackage{graphicx}

\usepackage{xcolor}

\title{No Stupid Questions: An Analysis of Question Query Generation for Citation Recommendation}

\author{
 \textbf{Brian D. Zimmerman\textsuperscript{1}},
 \textbf{Julien Aubert-Béduchaud\textsuperscript{2}},
 \textbf{Florian Boudin\textsuperscript{2}},
\\
 \textbf{Akiko Aizawa\textsuperscript{3}},
 \textbf{Olga Vechtomova\textsuperscript{1}}
\\
\\
 \textsuperscript{1}University of Waterloo,
 \textsuperscript{2}Nantes Université,
 \textsuperscript{3}National Institute of Informatics
\\
 \small{
   \textbf{Correspondence:} \href{mailto:bdzimmer@uwaterloo.ca}{bdzimmer@uwaterloo.ca}
 }
}

\begin{document}
\maketitle
\begin{abstract}
Existing techniques for citation recommendation are constrained by their adherence to article contents and metadata. 
We leverage GPT-4o-mini's latent expertise as an inquisitive assistant by instructing it to ask questions which, when answered, could expose new insights about an excerpt from a scientific article. 
We evaluate the utility of these questions as retrieval queries, measuring their effectiveness in retrieving and ranking masked target documents.
In some cases, generated questions ended up being better queries than extractive keyword queries generated by the same model. 
We additionally propose MMR-RBO, a variation of Maximal Marginal Relevance (MMR) using Rank-Biased Overlap (RBO) to identify which questions will perform competitively with the keyword baseline. 
As all question queries yield unique result sets, we contend that there are \textit{no stupid questions}.

\end{abstract}

\section{Introduction}
With the publication volume of scientific articles increasing rapidly, tools for enhancing researcher productivity have become invaluable. Scientists not only design and run experiments, but are also responsible for identifying and synthesizing related work to position their own contributions within community knowledge. The proliferation of Large Language Models (LLMs) has enabled scientists to expedite this workflow, enhancing productivity in literature surveyal, experimental design, data analysis, and writing research papers~\citep{10.1145/3487553.3527147,huang2023role, lu2024ai}. Although the delegation of these tasks is the most common role for LLMs in science, they can also catalyze scientific workflows in other ways.

\begin{figure}[ht]
    \centering
    \includegraphics[width=0.7\columnwidth]{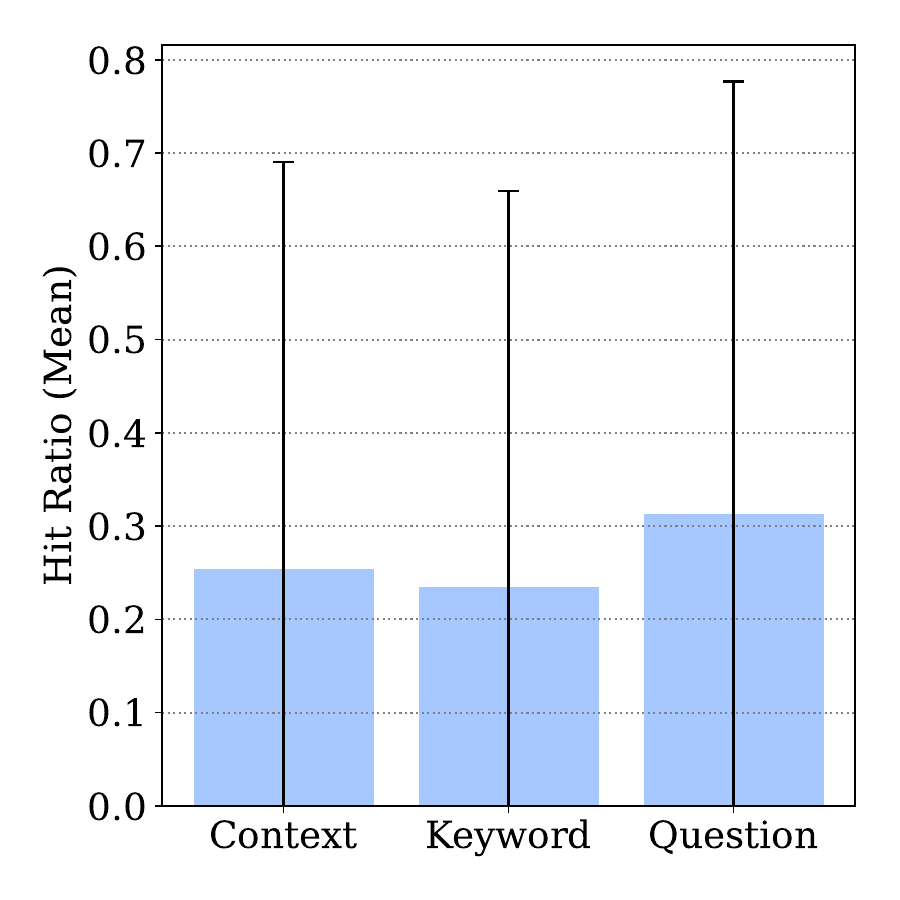}
    \caption{Context, Keyword (extractive), and the best-performing question from each group were evaluated as retrieval queries. Hit ratio reflects how often the target article appeared in the top 50, though predicting which question will succeed at inference time remains difficult.}
    \label{fig:hit-ratio}
\end{figure}


Citation recommendation is the task of recommending scientific papers for a citing motivation~\citep{ma2020review}. 
As a retrieval problem, citation recommendation is complex and exploratory in nature with no bound on the number of relevant documents.
Concretely, highly recommended citations should reflect both the citing motivation and the narrative that the researcher is trying to construct. 
In literature search, citation motivation should be formulated as a query. 
Queries are formed by either keyword extraction, bibliographic metadata, or a combination of each of them~\citep{bascur2023academic, martinez2025chatgpt, agarwal2025litllmtoolkitscientificliterature}.
Forming queries in this way constrains the narrative of the researcher to what has already been written. 

We position LLMs as inquisitive assistants to assist in writing scientific articles without imposing rigid constraints on an underlying narrative. 
LLMs have exhibited the ability to retain factual knowledge from training data. 
By harnessing this latent knowledge, we generate questions about research articles. 
Questions, unlike claims, have no truth value while also semantically conveying inquisitive intent. 
We conduct an analysis of the produced questions as search queries and explore techniques for identifying features of ``good questions''. 
 

\begin{table*}[ht]
    \centering
    {\small
\renewcommand{\tabularxcolumn}[1]{>{\centering\arraybackslash}m{#1}} 
\begin{tabularx}{\textwidth}{>{\centering\arraybackslash}m{2.5cm} >{\centering\arraybackslash}m{1.5cm} X}
    \hline
    \textbf{Query Type} & \textbf{Position} & \textbf{Query} \\
    \hline
    Keyword & 14 & "annotators identity", "data curation", "diversity perspectives", "conversational safety", "sociodemographic groups" \\
    \hline
    Question & 1 & What methodologies have been employed in previous research to assess the influence of annotators' backgrounds on their interpretations? \\
    \hline
    Question & 20 & What evidence exists to support the claim that diverse perspectives improve the robustness of datasets? \\
    \hline
    \hline
    \multicolumn{2}{>{\centering\arraybackslash}m{4cm}}{\textbf{Target Article}} & \textit{When Do Annotator Demographics Matter? Measuring the Influence of Annotator Demographics with the POPQUORN Dataset} \\
    \hline
\end{tabularx}
}
    \caption{Extractive keyword queries are limited by source material from which they were generated. Asking questions enables the model to use its expertise to seek articles that it may not otherwise have found.}
    \label{tab:reranked-positions}
\end{table*}

Our primary contribution is an analysis of questions generated by \texttt{GPT-4o-mini}\footnote{Specifically, we use \texttt{GPT-4o-mini-2024-07-18}} from masked excerpts of scientific articles.
These questions are analyzed on their utility as search queries, with some outperforming an extractive keyword baseline (Table~\ref{tab:reranked-positions}).
Although such questions may also function as valuable prompts for exploration and reflection to the researcher, in this work we focus on evaluating them by their utility as search queries.

We found that \texttt{GPT-4o-mini} was capable of producing questions which outperform the extractive baseline, both during retrieval and reranking (Figure~\ref{fig:hit-ratio}). 
At inference time, when the target article is unknown, it is often difficult to tell which questions from a batch are the most useful as queries. 

We propose the MMR-RBO scoring function for selecting questions which are both similar in utility to a verifiably good query while simultaneously distinct enough from other questions in the same batch. MMR-RBO is an extension of Maximal Marginal Relevance~\citep{carbonell1998use}, using Rank-Biased Overlap~\citep{webber2010similarity} as a similarity function. This method enables us to evaluate keyword queries and semantic queries on the basis of their retrieved results.

%
Our results show that all questions contribute uniquely—supporting the idea that there are \textit{no stupid questions}.



\begin{figure*}
    \centering
    \includegraphics[width=0.85\linewidth]{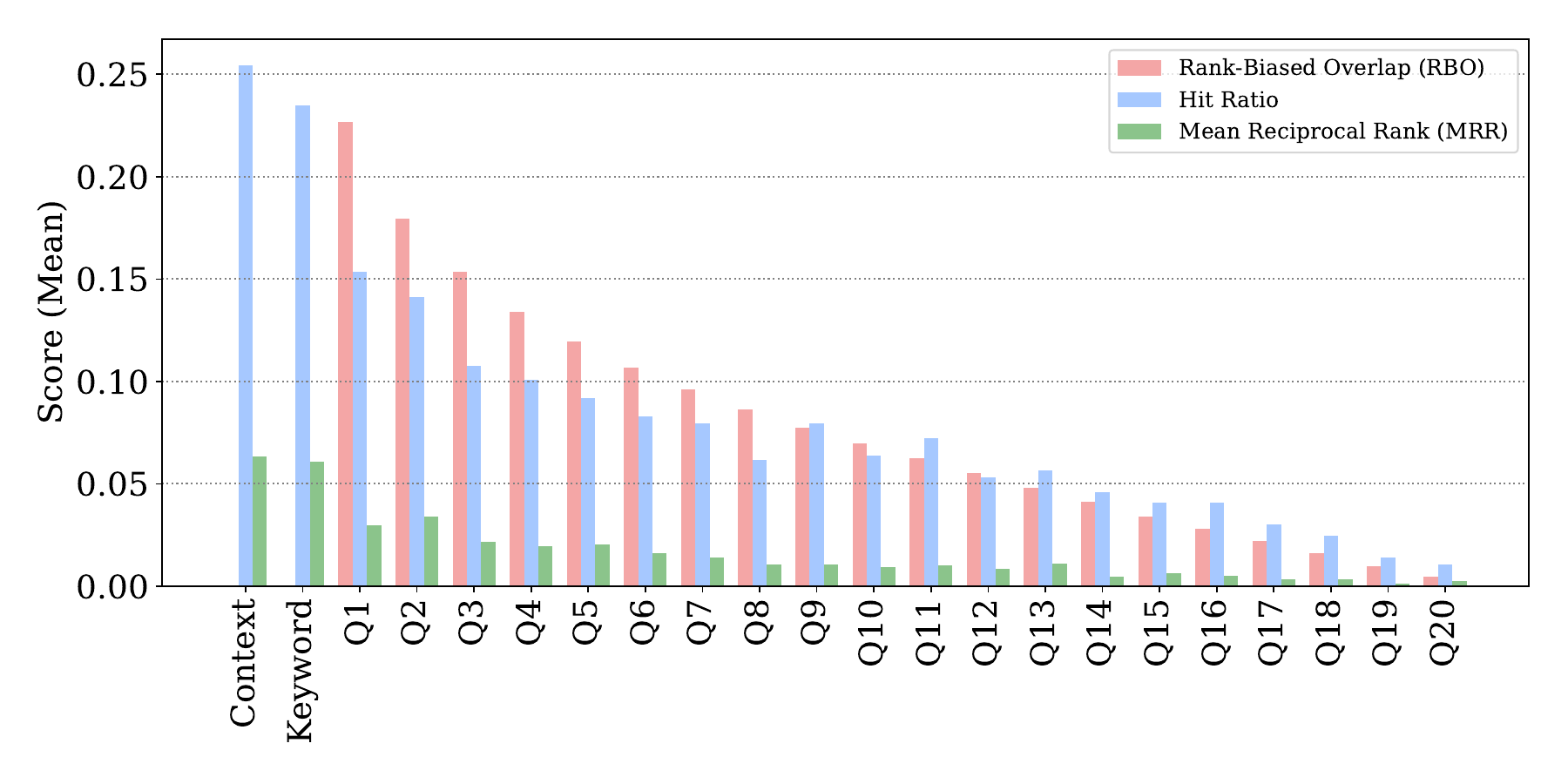}
    \caption{We compare context, keywords, and questions on their utility as retrieval queries. Questions are sorted by the Rank-Biased Overlap (RBO) between their retrieved documents and the documents retrieved by the context query. Higher RBO indicates a result set that is more similar to the result set of the context query.}
    \label{fig:rbo-ranked-retrieval}
\end{figure*}

\section{Experimental Design}
We downloaded all 1268 articles from the main track of EMNLP 2024 using the ACL Anthology to extract the related work sections from.

\subsection{Creating a Dataset of NLP Articles} 
We use GROBID~\citep{lopez2009grobid}, a tool for parsing scholarly articles, to extract both full text and citation spans from each article. Citation spans within the related work section of each article are replaced with a special mask token. We then split the related work section into paragraphs, forming our initial set of datapoints. We further disambiguate citation targets by matching ACL paper titles, allowing us to distinguish between citations to papers within the ACL Anthology and those external to it.

We select a candidate citation from each paragraph on the basis that it cites exactly one article and that article can be found within the ACL Anthology, our choice of retrieval corpus. We focused on citations with exactly one cited article to avoid confounding signals being encoded into generated questions. A total of 566 paragraphs met our filtering criterion.

We prompt \texttt{GPT-4o-mini} with the context around the candidate citation, asking for questions which could help identify articles to cite and strengthen the passage. We separately ask \texttt{GPT-4o-mini} to produce a query containing 5 keywords found within the context to serve as our baseline.





\subsection{Building a Retrieval Pipeline}
We use 94885 articles from ACL Anthology as our retrieval corpus. We encode titles and abstracts from the selected articles using SPECTER-2~\citep{Singh2022SciRepEvalAM} and store them within a FAISS index~\citep{johnson2019billion}. We encode each query with the SPECTER-2 ad-hoc query adapter and recorded the top 50 results when queried against the index. For each paragraph, we use the context, extracted keywords, and 20 questions as retrieval queries.

As result sets from the context queries most frequently contained the target document, we use context query results as the baseline for our reranking experiment. A total of 144 ($\sim{25\%}$) of the retrieval result sets contained the target document. We used a MS MARCO pre-trained MiniLM-v2 cross-encoder from Sentence Transformers~\citep{reimers2019sentence} to jointly encode title and abstract of retrieved documents against context, keyword, and question queries, producing a score by which they were reranked.

\subsection{Measuring Question Utility}

Our evaluation is contingent on two metrics: Mean Reciprocal Rank (MRR) (Eq.~\ref{eq:mrr}) and Rank-Biased Overlap (RBO)~\citep{webber2010similarity}. We consider MRR to be a measure of true query utility which indicates, when maximized, that the target document is ranked in the highest position. Because we don't know the target document at inference time, we examined several variables in search of a positive monotonic relationship with MRR.

\begin{equation}
\text{MRR} = \frac{1}{|Q|} \sum_{i=1}^{|Q|} \frac{1}{\text{r}_i}
\label{eq:mrr}
\end{equation}

RBO (Eq.~\ref{eq:rbo}) evaluates two result sets of size k, and returns a score based on their common elements and ordering. If we have a query $Q$ which is verifiably good, such as the extractive keyword baseline, we can compare it with a generated question $D_i$ using RBO (Eq.~\ref{eq:simq}). Due to the rank-weighted nature of RBO, two queries $D_i$ and $D_j$ could have low RBO between each other (Eq.~\ref{eq:simd}), while each having moderate RBO with $Q$. 
\begin{equation}
\text{RBO}(S, T, p, k) = \frac{X_k}{k} \cdot p^k + \frac{1 - p}{p} \sum_{d=1}^{k} \frac{X_d}{d} \cdot p^d
\label{eq:rbo}
\end{equation}

\begin{align}
\text{sim}_Q &= \text{RBO}(D_i, Q) \label{eq:simq} \\
\text{sim}_D &= \max_{D_j \in S} \text{RBO}(D_i, D_j) \label{eq:simd}
\end{align}

Maximal Marginal Relevance (MMR)~\citep{carbonell1998use}, is a $\lambda$-weighted combination of similarity functions used to find relevant, yet distinct documents in a collection. MMR-RBO ~(Eq.~\ref{eq:mmr-rbo}) is our variation of MMR, using RBO as a similarity function to find unique and high utility question queries. We evaluate the relationship between MRR and MMR-RBO.

\begin{align}
\text{MMR-RBO}(D_i) &= \lambda \cdot \text{sim}_Q - (1 - \lambda) \cdot \text{sim}_D \label{eq:mmr-rbo}
\end{align}

\section{Discussion}
\begin{figure}[h]
    \centering
    \includegraphics[width=0.90\columnwidth]{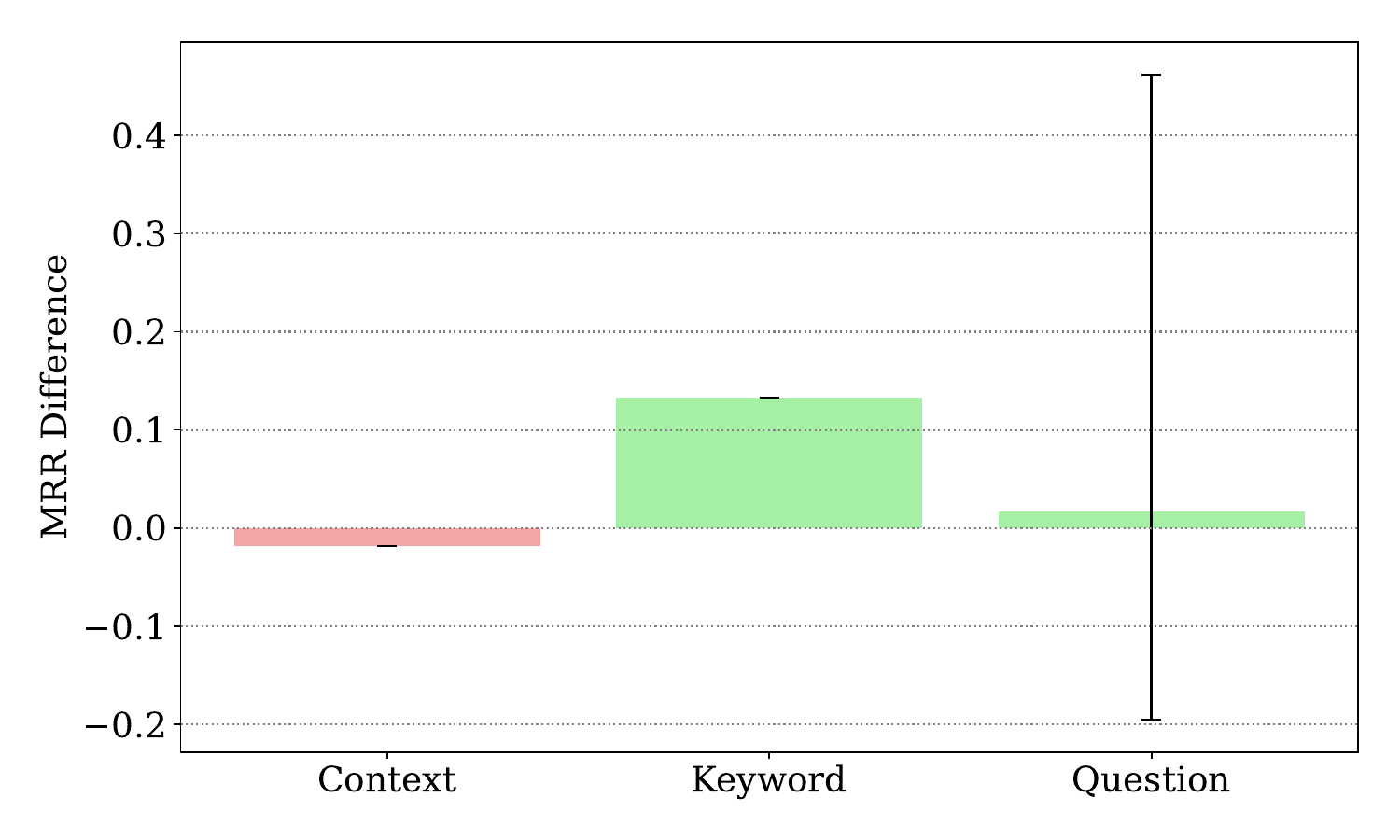}
    \caption{Difference in MRR by reranker query over the result set from the retriever. Colored bars are average across all paragraphs. Wicks represent minimum and maximum questions for each paragraph.}
    \label{fig:mrr-rerank}
\end{figure}

Finding the best question in a batch is a non-trivial process because the relationship between query and target is often unclear. Around $32\%$ of question batches contain at least one question with the target document in its retrieved results (Figure~\ref{fig:hit-ratio}), while Figure \ref{fig:rbo-ranked-retrieval} suggests that the question with the highest RBO with the context retrieves the target document around only $15\%$ of the time. It's a similar case with reranking where the extractive baseline outperforms questions, \textit{on average}, while the best question in each batch is a far better query with respect to the rank of the target document (Figure~\ref{fig:mrr-rerank}). This showcases the potential of questions as queries, but the difficulty remains in identifying the best one.

\subsection{The Needle in the Haystack}
\label{sec:needle-in-the-haystack}
\begin{figure}[h]
    \centering
    \includegraphics[width=0.90\columnwidth]{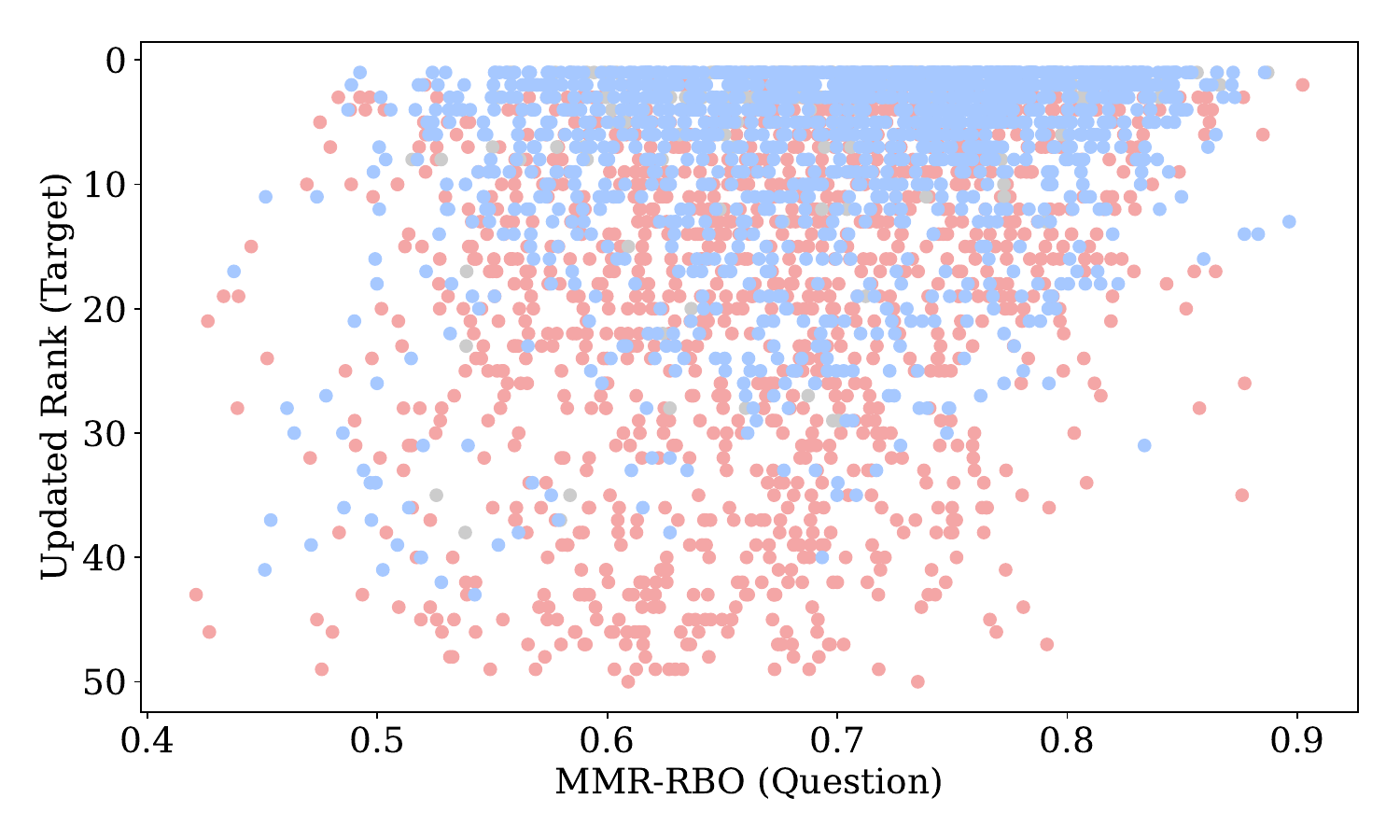}
    \caption{The observed correlation between MMR-RBO and article rank in question queries. Blue indicates an increase in rank, while red indicates a decrease in rank.}
    \label{fig:rerank-corr}
\end{figure}

In reranking, MMR-RBO can provide limited assistance in identifying the best query in a batch, but the parameter $\lambda$ proves extremely sensitive to the number of questions being generated. We observe a positive monotonic relationship between MRR and MMR-RBO (Figure~\ref{fig:rerank-corr}) by evaluating the average Fisher-transformed Spearman correlation across question batches (Table~\ref{tab:correlation}). As the number of questions per batch increases, the relationship between these values becomes significant across more batches as well. This suggests that, while there is an observed positive correlation between variables, MMR-RBO is highly susceptible to variance in small datasets. Generating more questions makes it easier to identify which questions outperform the extractive baseline.

\begin{table}[ht]
\centering
\begin{tabular}{cccc}
\textbf{Questions} & $\boldsymbol{\lambda}$ & \textbf{Batch Ratio} & \textbf{Spearman $\rho$} \\
\hline\hline
5  & 0.0 &0.0921&-0.1182\\
5  & 0.5 &0.0992&0.3273\\
5  & 1.0 &0.0851&0.0928\\
\hline
10 & 0.0 &0.1748&-0.1193\\
10 & 0.5 &0.1888&0.0255\\
10 & 1.0 &0.2027&0.1382\\
\hline
20 & 0.0 &0.2500&-0.1059\\
20 & 0.5 &0.2847&0.0515\\
20 & 1.0 &0.3680&0.1552\\
\hline\hline
20 & 0.0 &$\ast$&-0.2428\\
20 & 0.5 &$\ast$&0.1337\\
20 & 1.0 &$\ast$&0.2939\\
\end{tabular}
\caption{We evaluate the relationship between MRR and MMR-RBO on reranker queries by computing the mean Fisher-transformed Spearman correlation. Question batches were evaluated independently with a varying $\lambda$ hyperparameter for the MMR-RBO function. Batch ratio is the ratio of question batches with p-value < 0.05. Global Spearman correlation is denoted with $\ast$.}
\label{tab:correlation}
\end{table}

\subsection{Next Steps}
If questions are advantageous to keyword queries in that they are formed as natural language and carry semantic interpretability, they would be a more natural way for researchers to interface with retrieval systems. They should be further subjected to semantic evaluation, both by humans and using automated metrics, to validate their utility. We discuss one way in which a language model can be used to evaluate question quality in Appendix~\ref{sec:automated-evaluation} but a more thorough assessment could be formed as a document question answering task. Linking questions to claims made within a target document could be a future direction for scientific article retrieval. \citet{ajith2024litsearch} attempt exactly this by creating question dataset from ACL papers using GPT-4 and article authors. Curation of quality question data is an essential next step in bolstering the robustness of scientific retrieval models to question queries and it would enable researchers to more easily use their own questions to identify articles they’re seeking.

\section{Conclusion}
We generate questions using GPT-4o-mini from article excerpts and evaluated their utility as search queries using the MMR-RBO metric. 
We identify questions which, in some cases, outperformed an extractive keyword baseline in both retrieval and reranking while also observing a trend between MMR-RBO and the final rank of target documents. 
We also discuss several directions in which question query research for scientific document retrieval can be explored further.

\newpage
\section*{Limitations}
\subsection*{Automatic Generation of the Dataset}
A notable limitation of our dataset lies in its reliance on GROBID for  content and metadata extraction. Although GROBID is a popular tool for parsing structured bibliographic data from scientific PDFs, it is not infallible. The extraction process may introduce inaccuracies such as citation segmentation, misidentified section headers, and malformed reference entries. These errors can propagate through the dataset and negatively affect the performance and reliability of downstream tasks.
We explored Nougat~\cite{blecher2023nougat} for content extraction coupled to regular expression for citation identification. However, GROBID ultimately proved more suitable for our pipeline due to its ease of integration, reproducibility, and efficient processing times, even though it may not always produce fully accurate content.


\subsection*{MMR-RBO Variance Issues on Small Datasets}
The largest version of our dataset only contained 20 questions per batch. As mentioned in Section ~\ref{sec:needle-in-the-haystack}, the number of statistically significant batches in support of correlation between MMR-RBO and target document rank increases as the number of questions in a batch increases. This indicates that MMR-RBO is vulnerable to noise in high variance, small datasets. MMR-RBO as a scoring function for identifying good question queries will operate in a more stable manner against larger pools of questions.

\section*{Ethical Considerations}
\subsection*{Usage of LLMs in Writing Scholarly Articles}
Leveraging LLMs to assist with writing  scholarly articles should be approached with caution to prevent the inclusion of offensive content or the spread of misinformation. While LLMs have the potential to enhance learning by offering diverse perspectives and personalized experiences, the reliability of their outputs depends on the quality of the data they are trained on. 
If LLM-generated questions were applied in real-world scenarios, it would be crucial to address these concerns by reinforcing safety measures and actively ensuring that no misinformation is spread—thereby maintaining an accurate, ethical, and beneficial learning environment.

\subsection*{Reproducibility of Experiments}
We provide detailed descriptions of our methodology and will make our code and data publicly available under MIT license after publication.

\bibliography{custom}

\appendix

\section{Additional Evaluation}
\label{sec:automated-evaluation}

\subsection{Scoring Question Relevance using LLMs}

To estimate the topical relevance of a question with respect to a given set of keywords, we implemented a prompt-based method that leverages \texttt{Llama 3.1 8B}\footnote{\url{https://huggingface.co/meta-llama/Llama-3.1-8B}} as a probabilistic evaluator. The task is framed as a language modeling problem, where the model is prompted to assign a relevance score on a fixed scale from 1 (least relevant) to 5 (most relevant). The prompt is formulated in natural language as follows:

\begin{quote}
On a scale from 1 (worst) to 5 (best), score how relevant the question is to the topic characterized by the list of keywords.\\
\textbf{Question:} \texttt{\{question\}}\\
\textbf{Keywords:} \texttt{\{keywords\}}\\
\textbf{Answer:}
\end{quote}

The prompt is first tokenized using the language model’s tokenizer to generate input token IDs for inference. These tokens are then passed through the model in inference mode, and the logits corresponding to the next-token prediction are extracted. To narrow the output space, we limit evaluation to tokens corresponding to the numerical scores ``1'' through ``5''. Only those scores encoded as a single token are retained to ensure clear and unambiguous mapping. A softmax function is then applied to the filtered logits to obtain a probability distribution over the five candidate scores. This process yields a dictionary that maps each score to its corresponding model-assigned probability:
\begin{quote}
\texttt{\{"1": $p_1$, "2": $p_2$, ..., "5": $p_5$\}}
\end{quote}

Rather than relying solely on the top-ranked prediction, we compute a weighted score based on the full distribution. This allows us to account for the model’s uncertainty and provide a smoother evaluation signal. The score is calculated as the expected value of the predicted rating:

\[
 \text{Score} = \sum_{i=1}^{5} p_i \cdot i
\]

\begin{figure}[hb]
    \centering
    \includegraphics[width=0.9\linewidth]{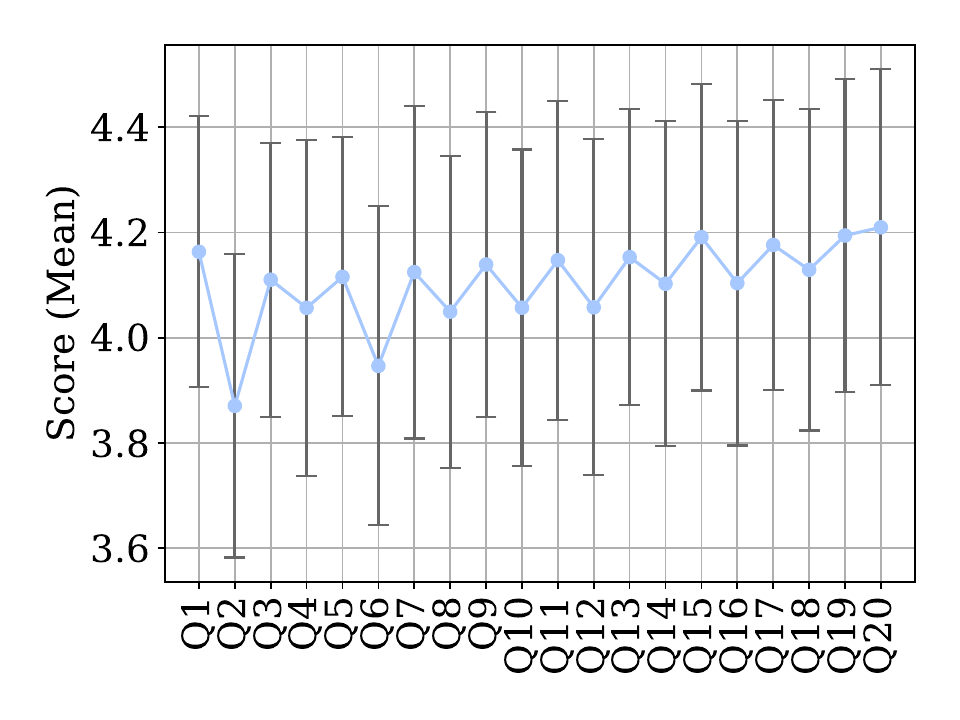}
    \caption{Although questions are scored by Llama individually, we observed an increasing trend when observing question scores in the order in which the questions were generated.}
    \label{fig:llm-scoring}
\end{figure}

We conduct this evaluation on 20 questions generated by \texttt{GPT-4o-mini}. As shown in Figure~\ref{fig:llm-scoring}, the quality of generated questions correlates with both their position in the sequence and the total number of questions. Notably, later questions tend to receive higher scores. This pattern suggests that as generation progresses, the model may benefit from contextual cues or internal calibration, resulting in questions that are increasingly aligned with the article's keywords. Whether these questions would be genuinely useful to a human remains an open question and requires further evaluation.

\end{document}